\documentclass[aps,prb,twocolumn,superscriptaddress,showpacs]{revtex4}
\usepackage{amsmath}
\usepackage{amsthm}
\usepackage{amssymb}
\usepackage{graphicx}
\usepackage{tabularx}
\usepackage[squaren]{SIunits}

\usepackage{color}

\newcommand{\bea}{\begin{eqnarray}}
\newcommand{\eea}{\end{eqnarray}}
\newcommand{\beq}{\begin{equation}}
\newcommand{\eeq}{\end{equation}}
\newcommand{\benu}{\begin{enumerate}}
\newcommand{\enu}{\end{enumerate}}

\newcommand{\TK}{T_{\rm K}}

\begin{document}
\title{
Kondo effect on the surface of 3D topological insulators:\\
Signatures in scanning tunneling spectroscopy
}
\date{\today}

\author{Andrew Mitchell}
\affiliation{Institut f\"ur Theoretische Physik, Universit\"at zu K\"oln, Z\"ulpicher
Stra\ss e 77, 50937 K\"oln, Germany}
\author{Dirk Schuricht}
\affiliation{Institute for Theory of Statistical Physics and JARA-Fundamentals of Future Information
Technology, RWTH Aachen University, 52056 Aachen, Germany}
\author{Matthias Vojta}
\affiliation{Institut f\"ur Theoretische Physik, Technische Universit\"at Dresden, 01062 Dresden, Germany}
\author{Lars Fritz}
\affiliation{Institut f\"ur Theoretische Physik, Universit\"at zu K\"oln, Z\"ulpicher
Stra\ss e 77, 50937 K\"oln, Germany}

\begin{abstract}
We investigate the scattering off dilute magnetic impurities placed on the surface of
three-dimensional topological insulators. In the low-temperature limit, the impurity
moments are Kondo-screened by the surface-state electrons, despite their exotic locking of
spin and momentum.
We determine signatures of the Kondo effect appearing in quasiparticle interference (QPI)
patterns as recorded by scanning tunneling spectroscopy, taking into account the
full energy dependence of the T matrix as well as the hexagonal warping of the surface Dirac
cones. We identify a universal energy dependence of the QPI signal at low scanning
energies as the fingerprint of Kondo physics, markedly different from the signal due to
non-magnetic or static magnetic impurities.
Finally, we discuss our results in the context of recent experimental data.
\end{abstract}

\maketitle


\section{Introduction}

Topological insulators (TIs) in both two and three spatial dimensions constitute an
active topic of current condensed matter
research.\cite{KM,BHZ,koenig,FKM,MooreBalents,QiHughesZhang,roy} The non-trivial bulk
band topology of three-dimensional (3D) strong topological insulators causes the crossing
of surface states at time-reversal invariant points in the surface Brillouin zone and
gives rise to a two-dimensional surface metal. In the vicinity of such crossing
points, the effective surface theory takes the form of a Dirac equation of massless
fermions, where spin and momentum are locked together.

A fundamental property of this ``helical'' surface metal is a suppression of
backscattering: Electrons with opposite momenta have orthogonal spin projections,
such that impurity scattering $\vec{k} \leftrightarrow -\vec{k}$ is impossible without a 
spin flip. As a result, the metallic state is protected from the influence of
non-magnetic disorder, and weak localization is replaced by weak antilocalization.\cite{beenakker07,ryu07}
This scenario of forbidden backscattering has been tested in recent
experiments\cite{yazdani,kapitulnik,xue,madhavan,hsieh,xia} utilizing powerful
Fourier-transform scanning tunneling spectroscopy\cite{eigler,davis} (FTSTS). In this
technique, energy-dependent spatial variations of the local density of states (LDOS) are
analyzed in terms of quasiparticle interference (QPI), i.e., quasiparticle scattering
processes due to impurities. The QPI results obtained on 3d TIs such as Bi$_{1-x}$Sb$_x$
and Bi$_2$Te$_3$ were found to be consistent with a heuristic picture of electron
scattering in a helical liquid, with backscattering being suppressed.

These results prompt the question as to how scattering from \emph{magnetic} impurities on the
surface of TIs is manifest in observables such as the QPI patterns obtained by FTSTS. In
fact, recent experiments\cite{madhavan} on Bi$_2$Te$_3$ doped with dilute magnetic Fe
atoms purport to demonstrate from the QPI pattern signatures of time-reversal symmetry
breaking.
However, one must be careful to distinguish a fluctuating magnetic moment
from one which is static on the large timescale of the STS experiment. The latter
situation implies magnetic long-range order, whose existence requires a sufficient
density of magnetic moments and low temperature. Then, every impurity moment is
polarized, and time-reversal symmetry is broken.\cite{xu}
Interestingly, it has been shown theoretically that such static magnetic impurities do
not lead to backscattering being visible in QPI; within lowest-order Born approximation,
a static local field is entirely invisible in QPI.\cite{zhou09,guofranz}

In this paper, we focus instead on the case of fluctuating magnetic impurities,
relevant to the dilute limit. The interaction between the impurity moment and the
electrons of the surface metal leads to mutual spin flips, such that backscattering could
be allowed although time-reversal symmetry remains unbroken. It is such
spin-flip processes which lead to Kondo screening of the impurity moment in standard
metals.\cite{hewson}
Therefore the key question, also relevant to the experiments of
Ref.~\onlinecite{madhavan}, pertains to the signatures in QPI of the Kondo interaction
between the helical metal and the impurity.
To answer this, we solve the problem of a single Kondo impurity on the surface of a
3D TI numerically exactly, and calculate the induced QPI pattern which of course now includes
inelastic scattering off the magnetic moment.

Our main findings are as follows.
(i) The magnetic impurity is described by a standard SU(2)-symmetric impurity model,
despite spin--momentum locking and hexagonal warping effects of the surface states of a
real TI. As a result, the impurity moment is always Kondo screened in the low-temperature
limit, unless the chemical potential is tuned exactly to the Dirac point.
(ii) While scattering off a Kondo impurity does not open new scattering channels in
momentum space as compared to a non-magnetic impurity, it leads to a distinct {\em
energy} dependence of the QPI pattern, which moreover exhibits universal scaling in terms
of both scanning energy and temperature. A strong enhancement of the QPI intensity near
the Fermi level is therefore a signature of scattering caused by fluctuating magnetic
impurities.

The body of the paper is organized as follows. We start by introducing the model and methods in
Sec.~\ref{sec:model}. The Kondo effect on the surface of 3D TIs is discussed in
Sec.~\ref{sec:Kondo}. Sec.~\ref{sec:magimp} is then devoted to the QPI patterns from
Kondo impurities, with an emphasis placed on universal features. The QPI signal from
non-magnetic impurities is shown for comparison in Appendix \ref{sec:scalar} while we briefly discuss a static magnetic impurity in Appendix~\ref{sec:staticmag}.
The implications of our results for experimental data are discussed in the concluding
section~\ref{sec:conclusion}.

We note that the Kondo effect on the surface of 3D TIs was discussed before in
Refs.~\onlinecite{zitko,kiseok}, but without taking into account hexagonal warping and
without a discussion of QPI. QPI patterns for surfaces of 3D TIs have been calculated for
different types of impurities in Refs.~\onlinecite{guofranz,zhang,yazdani,madhavan,zhou09}, but
no link to Kondo physics was made. Very recently, inelastic scattering from excited
states of magnetic impurities was discussed in Ref.~\onlinecite{thalmeier}, leading to
features at elevated-energy in tunneling spectra.


\section{Model and Methods}\label{sec:model}

\subsection{Effective surface metal}

Surface states of 3D TIs are described by an effective Dirac theory. However, such a
linearized model applies only in the immediate vicinity of the crossing point of the
surface bands, while lattice effects must be taken into account at higher energies. For
Bi$_2$Te$_3$ this results in a breaking of the continuous rotation symmetry around the
Dirac point down to $C_{3v}$, leading to so-called hexagonal warping of the iso-energy
contours,\cite{fu} as is seen experimentally.\cite{hasan2009,chen09,yazdani,madhavan}

The free Hamiltonian of the surface metal reads\cite{fu,zhang}
\begin{eqnarray}
H_0=\int d^2{\bf{k}} \left(\Psi^\dagger_{\bf{k}\uparrow}, \Psi^\dagger_{\bf{k}\downarrow}\right) \hat{\mathcal{H}}_{\bf{k}}\left( \begin{array}{c}   \Psi^{\phantom{\dagger}}_{{\bf{k}}\uparrow}  \\
\Psi^{\phantom{\dagger}}_{{\bf{k}}\downarrow} \end{array} \right)
\end{eqnarray}
where
\begin{eqnarray}\label{eq:H0}
\hat{\mathcal{H}}_{\bf{k}}=\hbar v_F \left[ ({\bf{k}}\times \boldsymbol{\sigma})\cdot {\bf{e}}_z
+ A^2 k^3 \cos(3\phi_{\bf{k}})  \sigma_z \right]-\mu \;.
\end{eqnarray}
Here $\boldsymbol{\sigma}$ is a vector of the Pauli matrices, $k=|\bf{k}|$ is the
magnitude of the momentum vector relative to the Dirac point, and
$\phi_{\bf{k}}=\tan^{-1}(k_y/k_x)$ is its azimuthal angle measured with respect to the
$\hat{x}$ axis. The $\Gamma$--$K$ direction thus corresponds to $\phi_{\bf{k}}=0$ while
$\Gamma$--$M$ corresponds to $\phi_{\bf{k}}=\pi/6$, following Ref.~[\onlinecite{zhang}].
The cubic term $\propto A^2$ accounts for hexagonal warping, and $\mu$ denotes the
chemical potential.

The spectrum of the above Hamiltonian (with $\hbar=1$ hereafter) is given by
\begin{eqnarray}\label{eq:spectrum}
E_{\pm}({\bf{k}})=\pm v_F \sqrt{k^2+\left[A^2 k^3 \cos \left(3\phi_{\bf{k}} \right)\right]^2}-\mu\;.
\end{eqnarray}
The free Green function,
$\hat{G}^{(0)}(\textbf{k},\omega)=[\omega+i0^+-\hat{\mathcal{H}}_{\bf{k}}]^{-1}$,
takes a diagonal form in the quasiparticle basis,
$\hat{\mathcal{G}}^{(0)}_{ab}(\textbf{k},\omega)=\delta_{ab}/[\omega+i0^+-E_a(\textbf{k})]$,
$a,b=\pm$.
The density of states (DOS) follows from
\begin{eqnarray}\label{eq:dosdef}
\rho^{(0)}(\omega)=-\frac{1}{\pi N}\text{Im}~\text{Tr}~\hat{\mathcal{G}}^{(0)}(\bf{k},\omega),
\end{eqnarray}
where the trace accounts for the sum over $\pm$ as well as $\bf{k}$, and $N$ is a
suitable normalization factor (equal to the number of $\bf{k}$ points).\cite{cont_note}
$\rho^{(0)}(\omega)$ is linear in $|\omega|$ at low energies around $\omega=-\mu$,
characteristic of massless Dirac fermions.

In the following we shall employ parameters $v_F/a_0=0.73\,$eV ($a_0=1$ is a lattice
constant acting) as our energy unit, and $A^2=2.23$ to make contact with
experiments.\cite{madhavan} For convenience we shall use $v_F/a_0$ as
high-energy cut-off for the conduction band, $\rho^{(0)}(\omega)\equiv
\rho^{(0)}(\omega) \theta(v_F-|\omega +\mu|)$ -- this is mainly needed to generate an
input for the numerical treatment of the impurity problem in Sec.~\ref{sec:Kondo}.
(For a 3D TI, a natural cutoff is set by the size of the bulk gap.)
Generically, the Fermi level is not at the Dirac point; for example, in
Ref.~\onlinecite{madhavan} the chemical potential is $\mu/v_F=0.137 \equiv 100$\,meV.
Below we consider this case explicitly, and also the special case where $\mu=0$, which is
potentially attainable, since surfaces of TIs can be individually gated.

\begin{figure}[!t]
\includegraphics[width=0.49\textwidth]{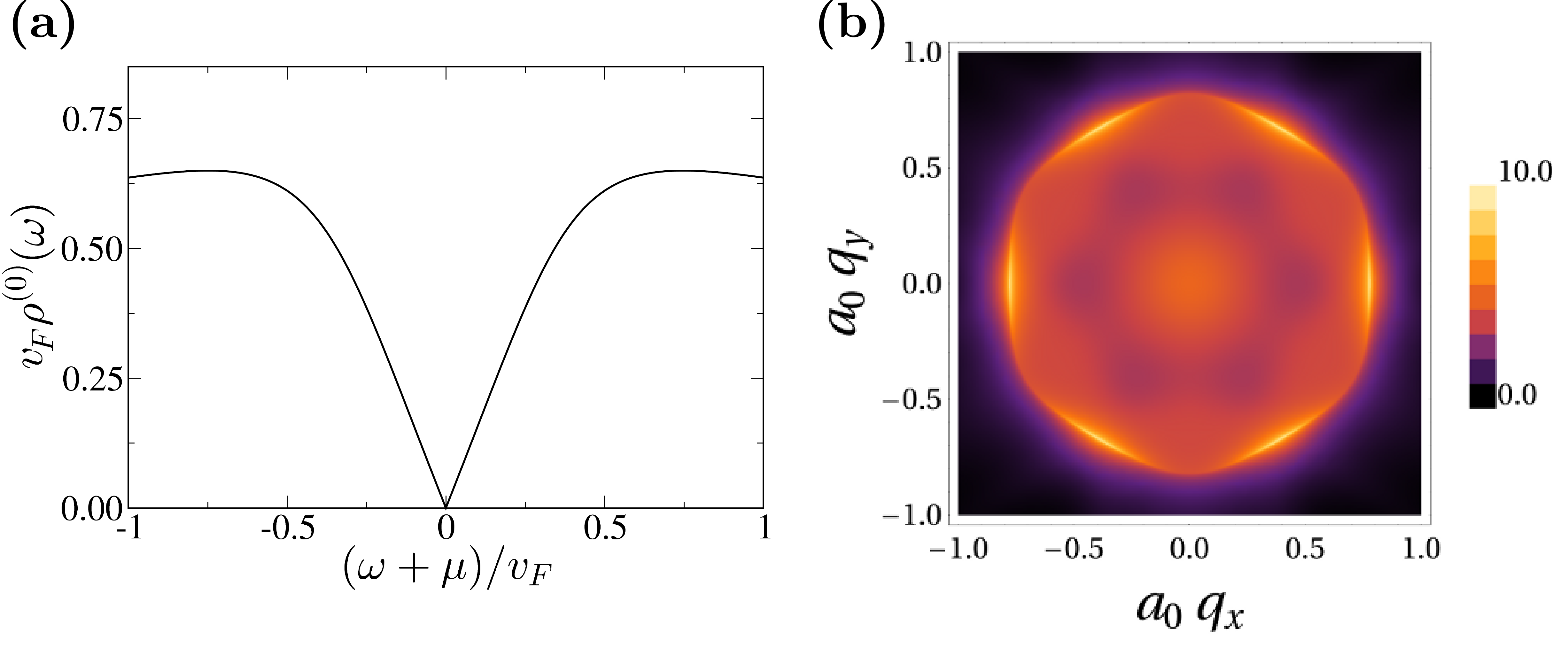}
\caption{\label{fig:dos}(color online)
(a) Density of states, $\rho^{(0)}(\omega)$, for the effective
surface theory, Eq.~(\ref{eq:H0}) in the absence of
impurities, for $A^2=2.23$. At energies around $\omega=-\mu$ one
obtains linear pseudogap behavior, which is modified at higher
energies due to hexagonal warping. (b) QPI pattern $\Delta
\rho({\bf{q}},\omega)$
for a TI with $\rho^{(0)}(\omega)$ as in (a), using
$T^0(\omega)=1$ appropriate for a potential scatterer in Born approximation. Plotted for
$\mu/v_F=0.137\equiv 100$\,meV and scanning energy
$\omega=200$\,meV.
}
\end{figure}

The local Green function on the surface of the TI has a matrix structure in spin space
which turns out to be \emph{diagonal}:
\begin{eqnarray}\label{eq:absenceoff}
\hat{G}^{(0)} \left(\textbf{r}=0, \omega_n\right)=\mathbb{I} f_{\omega_n,\mu}\;,
\end{eqnarray}
in terms of Matsubara frequency $\omega_n$. Here
\begin{eqnarray}\label{eq:lg}
f_{\omega_n,\mu}= \int_{\bf{k}} \frac{i\omega_n+\mu}{\left(i\omega_n+\mu \right)^2-v_F^2
  k^2-v_f^2 A^4 k^6 \cos \left(3\phi_{\bf{k}} \right)^2} \;,
\end{eqnarray}
where $\int_{\bf{k}}=\int \frac{d^2 {\bf{k}}}{(2\pi)^2}$, such that
$\rho^{(0)}(\omega) = -2 {\rm Im}\,f_{\omega,\mu} / \pi$.
As shown in Appendix~\ref{app:localgreen}, the off-diagonal elements of
$\hat{G}^{(0)}(\textbf{r}=0)$ vanish on angular integration.


\subsection{Impurities and QPI}

The FTSTS technique exploits scattering of charge carriers from impurities in an
otherwise translationally-invariant system.\cite{eigler,davis} To this end, a real-space
map of the tunneling conductance is recorded at fixed energy.
The Fourier transform of this map to
momentum space reveals characteristic wavevectors of LDOS inhomogeneities which can be
understood as energy-dependent Friedel oscillations (or QPI). In the simplest
approximation, these wavevectors correspond to scattering processes of
quasiparticles between different points of the dispersion iso-energy contour at the
scanning energy.\cite{ripples}

If scattering centers are dilute, it is sufficient to consider a single impurity. Its
effect is described by the T matrix, $\hat{T}_{\textbf{k},\textbf{k}'}(\omega)$, such
that the full electronic Green function reads
\begin{eqnarray}\label{eq:tmatrix}
\hat{G}(\textbf{k},\textbf{k}',\omega)&=&\hat{G}^{(0)}(\textbf{k},\omega)\delta_{\textbf{k},\textbf{k}'}
 \nonumber \\
 &+&\hat{G}^{(0)}(\textbf{k},\omega)\hat{T}_{\textbf{k},\textbf{k}'}(\omega)\hat{G}^{(0)}(\textbf{k}',\omega)\,,
\end{eqnarray}
which remains energy-diagonal in equilibrium/linear response.
The real-space LDOS is
\begin{equation}\label{Eq:deltarho}
\rho({\bf{r}},\omega)= -\frac{1}{\pi}{\rm{Im}}~
\int_{\bf q} e^{i{\bf q}\cdot\textbf{r}}~{\rm{Tr}}~\hat{G}(\textbf{k},{\bf{k}}-{\bf{q}},\omega)\,,
\end{equation}
where the trace again accounts for the sum over $\textbf{k}$ as well as
the spin components.
For isotropic scattering and an inversion-symmetric host, the Fourier transform
$\rho({\bf{q}},\omega)$ is real, and its impurity-induced piece is related to
the T matrix via
\begin{equation}\label{Eq:deltarho1}
\Delta \rho({\bf{q}},\omega)= -\frac{1}{\pi}{\rm{Im}}~ {\rm{Tr}}~\hat{G}^{(0)}
({\bf{k}},\omega) \hat{T}_{{\bf{k}},{\bf{k}}-{\bf{q}}}(\omega)  \hat{G}^{(0)}
({\bf{k}}-{\bf{q}},\omega).
\end{equation}

\emph{If} the T matrix is diagonal in spin space (see below), decomposition
using the Pauli matrices gives
\begin{eqnarray}\label{eq:g0}
\hat{T}_{{\bf{k}},{\bf{k}}'}(\omega)=T^0_{{\bf{k}},{\bf{k}}'}(\omega)
\mathbb{I}+T^1_{{\bf{k}},{\bf{k}}'}(\omega) \sigma_z\;,
\end{eqnarray}
with
$T^{0,1}_{{\bf{k}},{\bf{k}}'}(\omega) =
\left[T_{{\bf{k}},{\bf{k}}',\uparrow}(\omega) \pm T_{{\bf{k}},{\bf{k}}',\downarrow}(\omega)\right]/2$.
For TI surfaces in the absence of hexagonal warping it was shown in
Ref.~\onlinecite{guofranz} that only the part proportional to $\mathbb{I}$ leads to a
modulation of the spin-integrated LDOS in Eqs.~(\ref{Eq:deltarho},\ref{Eq:deltarho1}).
(The part proportional to $\sigma_z$ causes opposite modulations for both spin
directions.) Using symmetry properties, we have verified that this still holds for the
full model including hexagonal warping: the argument parallels that given in
Appendix~\ref{app:localgreen}.

Specializing further to a point-like impurity with
$\hat{T}_{{\bf{k}},{\bf{k}}'}(\omega) \equiv \hat{T}(\omega)/N$,
we have
\begin{equation}\label{eq:qpidef}
\Delta \rho({\bf{q}},\omega)=
-\frac{1}{\pi}{\rm{Im}}~T^0(\omega) ~ {\rm{Tr}}~
\hat{G}^{(0)}({\bf{k}},\omega)  \hat{G}^{(0)}({\bf{k}}-{\bf{q}},\omega)
\end{equation}
such that the momentum dependence of the QPI signal at fixed $\omega$ is completely
determined by $\hat{G}^{(0)}$.
A sample QPI image is displayed in Fig.~\ref{fig:dos} -- this is similar to published
results.\cite{zhou09} One clearly observes a breaking of the circular symmetry due to
hexagonal warping of the iso-energy contour at this energy.

As we show below, a (local) Kondo impurity does not modify the overall momentum
dependence of the QPI pattern, but will lead to a non-trivial energy dependence.


\subsection{Anderson impurity model}\label{sec:effaim}

To describe a dynamic magnetic scatterer, we consider Anderson's model for a point-like
correlated impurity,\cite{hewson} $H=H_0+H_{\text{imp}}+H_{\text{hyb}}$, with
\begin{equation}\label{Eq:Himp}
\begin{split}
 H_{\text{imp}}&=\epsilon_d (\hat{n}^d_{\uparrow}+\hat{n}^d_\downarrow)
  + U
 \hat{n}^d_{\uparrow}\hat{n}^d_\downarrow \;,\\
 H_{\text{hyb}}&=g  \sum_{\sigma}  d^{\dagger}_\sigma
   \Psi^{\phantom{\dagger}}_{\sigma}(r=0) +{\text{H.c.}}.
\end{split}
\end{equation}
Here $\Psi_{\sigma}(r\!=\!0)=\sum_{\textbf{k}}\Psi_{\textbf{k}\sigma} / \sqrt{N}$,
$\hat{n}^d_{\sigma}=d^{\dagger}_\sigma d^{\phantom{\dagger}}_\sigma$, and $\epsilon_d<0$
and $U>-\epsilon_d$ are the local level energy and Coulomb repulsion, respecively.

Considering the spin-momentum locking and the hexagonal warping of the TI surface
electrons in $H_0$, one might have expected an unconventional impurity problem. However,
a standard SU(2) spin-symmetric impurity problem is obtained, with the complexity of the
helical surface metal entering only through the unusual DOS,
Eq.~(\ref{eq:dosdef}).
The derivation of such a pseudogap Anderson (or Kondo) model has been established before
for impurities in $d$-wave superconductors\cite{FV05} as well as for TIs with a perfect
Dirac structure,\cite{zitko,kiseok} and we give here an efficient proof which also covers
hexagonal warping.
Rather than using a decomposition into angular modes as in
Refs.~\onlinecite{zitko,kiseok}, a more direct way is to use the path-integral
formulation. Since the conduction-electron bath is Gaussian, we integrate it out
exactly\cite{FV05} to derive a local retarded impurity problem. The local
non-interacting part of the action for the d-levels in Matsubara formalism reads
\begin{equation}\label{eq:action}
\mathcal{S}_0 =-T \sum_{ \omega_n} \overline{{\bf{d}}}^T \left[
  g^2\hat{G}^{(0)}\left(\textbf{r}=0,i\omega_n\right)-\mathbb{I}(\epsilon_d-i\omega_n) \right]
{\bf{d}}\;,
\end{equation}
with $\hat{G}^{(0)}\left(\textbf{r}=0,i\omega_n\right)$ given in
Eq.~(\ref{eq:absenceoff}) and ${\bf{d}}=(d_\uparrow,d_\downarrow)$. The resulting local
model is equivalent to the standard Anderson model because
$\hat{G}^{(0)}\left(\textbf{r}=0,i\omega_n\right)$ is diagonal. The hybridization
function characterizing this impurity problem is
\begin{eqnarray}\label{eq:hybfunc}
 \Delta (\omega)=-\text{Im}~g^2
 f_{\omega,\mu} = \frac{\pi}{2} g^2 \rho^{(0)}(\omega)\;.
\end{eqnarray}

Importantly, the impurity Green function $\hat{G}^d(\omega)$
is diagonal due to the absence of off-diagonal terms in the quadratic
action of the d-levels, Eq.~(\ref{eq:action}). In the absence of a magnetic field, we
thus have
\begin{eqnarray}\label{eq:gfaim}
\hat{G}^{d} (\omega)=\frac{1}{\omega+i0^{+}-\epsilon_d -g^2 f_{\omega,\mu}-\Sigma^{d}(\omega)}~\mathbb{I} \;.
\end{eqnarray}
Here $\Sigma^{d}(\omega)$ is the interaction part of the impurity
self-energy.
The corresponding T matrix is then
$\hat{T}_{\textbf{k},\textbf{k}'}(\omega)=g^2 \hat{G}^d (\omega)/N \propto \mathbb{I}$,
and one expects a non-trivial
response from Anderson-type impurities in QPI according to Eq.~\eqref{eq:qpidef}.


\section{Kondo effect}\label{sec:Kondo}

The Anderson impurity model, Eq.~(\ref{Eq:Himp}), can describe both the formation and the
subsequent screening of local magnetic moments. Charge fluctuations induce two broadened
peaks (``Hubbard satellites'') in the impurity spectral function at $\epsilon_d$ and
$(U+\epsilon_d)$. The physics at energies smaller than ${\rm
min}(-\epsilon_d,U+\epsilon_d)$ is dominated by spin-flip processes leading to Kondo
screening below a temperature $\TK$ -- a phenomenon which is sensitive to the
conduction-band DOS near the Fermi level.\cite{hewson}
Given the unusual DOS of the TI's surface metal, which vanishes linearly at $\omega=-\mu$,
the Kondo effect should be discussed separately in the  two cases: (A) the chemical
potential is tuned to the Dirac point, $\mu=0$; (B) finite chemical potential, $\mu\ne
0$.

We will obtain numerical results for the Anderson impurity model using Wilson's numerical
renormalization group (NRG) technique.\cite{nrg:rev} In the following, the hybridization
function, Eq.~(\ref{eq:hybfunc}), is discretized logarithmically using $\Lambda=3$, and
$\sim 3000$ states are retained at each step of the iterative diagonalization. The
results of $z=3$ interleaved calculations\cite{ztrick} are then combined for optimal
results. The full density matrix\cite{asbasis,fdmnrg} is calculated,  and from it the
impurity spectral function is determined numerically\cite{fdmnrg} as a full function of
energy, $\omega$, at arbitrary temperature, $T$.
The real part of $G^d$ is obtained using
Kramers-Kronig relations. The T matrix and hence QPI can then be calculated via
Eq.~(\ref{eq:qpidef}).

We note that the full DOS at the surface of a 3D TI will also have higher-energy
contributions from bulk states, not captured by our modelling. Consequently, our
calculation cannot establish a quantitative link between the parameters of the Anderson
model and $\TK$. At present, there is no experimental information available on the actual
values of $\TK$ for concrete TI materials and impurities.
Therefore we choose parameters of Eq.~(\ref{Eq:Himp}) such that $\TK$ attains values of
order 1~K.
For most calculations, we shall employ parameters $U/v_F=0.3$,
$\epsilon/v_F=-0.1$, and $g/v_F=0.168$ (chosen to match some elevated-energy features of
the data in Ref.~\onlinecite{madhavan}), corresponding to a moderately correlated
impurity. Results are shown for $T=0$ unless otherwise noted.


\subsection{Chemical potential at the Dirac point}\label{sec:psaim}

For $\mu=0$, the density of states at low energies is $\rho^{(0)}(\omega)\propto |\omega
|^r$ with $r=1$. The low-energy physics is thus that of the pseudogap Kondo
model.\cite{withoff,ingersent,logan,VF04,FV04}
For the case of particle--hole symmetry in both the impurity ($\epsilon_d=-U/2$) and the
bath, screening is absent for all parameters, and the impurity moment remains free
down to the lowest energy/temperature scales.
In contrast, if particle--hole symmetry is broken, a quantum phase transition occurs,
separating the local-moment and Kondo screened phases. However, the latter requires
strong particle--hole asymmetry and impurity-host coupling, such that screening is
less likely to occur at the Dirac point.

\begin{figure}[t]
\begin{center}
\includegraphics[width=0.47\textwidth]{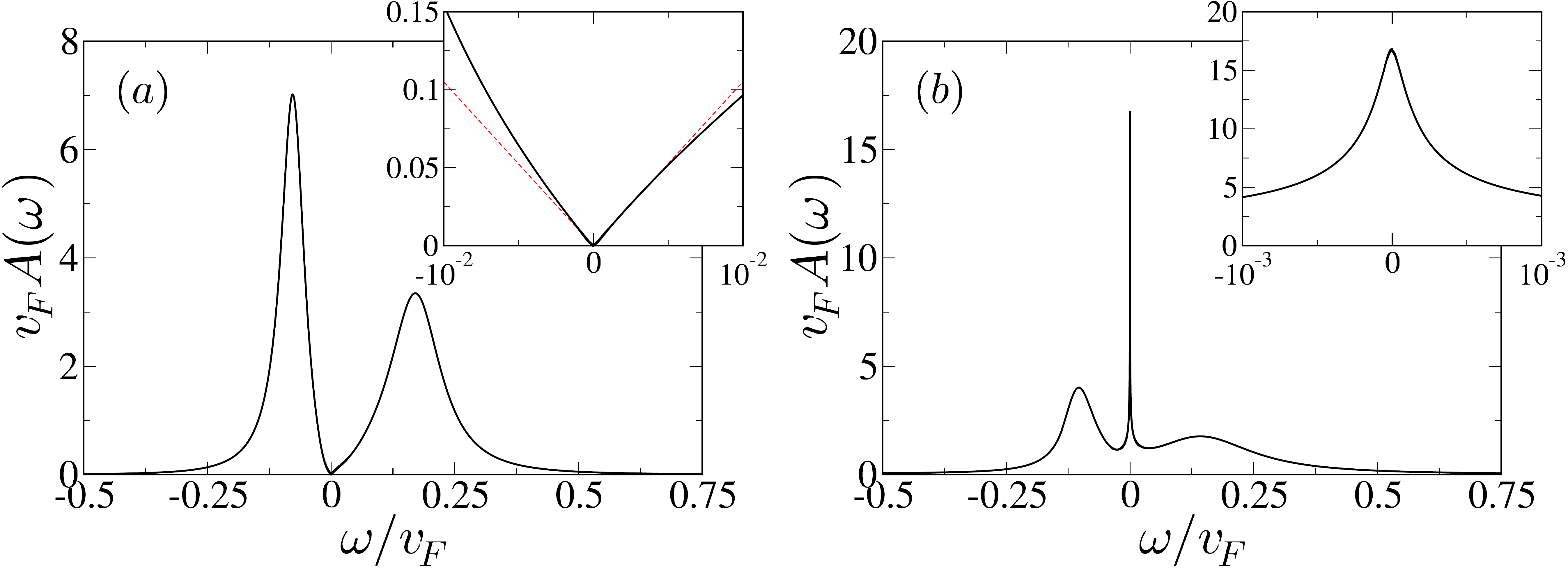}
\caption{\label{fig:Gimp_LM} (Color online)
Impurity spectral function $v_FA(\omega)$ vs energy $\omega/v_F$ at $T=0$
for $U/v_F=0.3$, $\epsilon/v_F=-0.1$, and $g/v_F=0.168$. (a) Deep in
the local moment phase for $\mu=0$. Inset shows a close-up in the
vicinity of the Fermi level, with the linear soft-gap in the free
density of states showing up at low energies (red dotted line is
$v_FA(\omega)\sim |\omega|$). (b) Deep in the Kondo screened phase,
$\mu=100$\,meV. Inset shows a close-up of the Kondo resonance.}
\end{center}
\end{figure}

A typical impurity spectral function
$A(\omega)=-\text{Im}~G_{\sigma}^d(\omega)/\pi$ for $\mu=0$ in the local-moment
phase at $T=0$ is shown in Fig.~\ref{fig:Gimp_LM}(a). At high energies
($\omega/v_F\approx -0.1$ and $\omega/v_F\approx 0.2$) clear signatures of the Hubbard
satellites are observed. At low energies (see inset) spectral weight is suppressed due to
the pseudogapped free density of states.\cite{logan,vb01}

\subsection{Chemical potential away from the Dirac point}\label{sec:fcp}

For finite chemical potential, $\mu\ne 0$, there is a finite density of
states at the Fermi level. The impurity is always
screened by the Kondo effect on the lowest energy scales (although the
Kondo temperature, $\TK$, itself might be very small).
Screening is reflected in the impurity spectral function by a narrow
resonance around the Fermi level. Together with the high-energy
Hubbard satellites, this three-peak structure is the classic hallmark
of the Kondo effect.

In Fig.~\ref{fig:Gimp_LM}(b) we plot the spectral function
for an impurity with for the same parameters as in panel (a), but with
$\mu=100$\,meV. The inset shows a close-up of the Kondo resonance, of
width $\TK\approx 10^{-4}v_F \sim 1$\,K [we define $\TK$ via
$A(\omega=\TK)=\tfrac{1}{2}A(\omega=0)$].

For $\mu$ values closer to the Dirac point, the impurity model will display non-trivial
crossover phenomena,\cite{vfb10} different from those of the standard Kondo problem,\cite{hewson} due
to the non-constant DOS and the proximity to the $\mu=0$ quantum phase
transition. Such crossovers can be expected when $\mu \lesssim\TK$ and are
not present in Fig.~\ref{fig:Gimp_LM}(b).


\section{QPI from dynamic magnetic impurities}\label{sec:magimp}

We now calculate the QPI pattern, $\Delta \rho({\bf{q}},\omega)$, induced
by a dynamic magnetic impurity on the surface of a 3D TI, first for the generic situation
of finite chemical potential and then for the special case where the chemical potential
is tuned to the Dirac point. We recall that $\Delta \rho({\bf{q}},\omega)$ is real for
our case of a single impurity; experiments typically extract its absolute value.


\subsection{Kondo phase}

A finite chemical potential implies Kondo screening at lowest temperatures.
We divide our analysis into the regimes of elevated and low energies.


\subsubsection{Elevated energies, $\omega \gg \TK$}

\begin{figure}[t]
\includegraphics[width=0.49\textwidth]{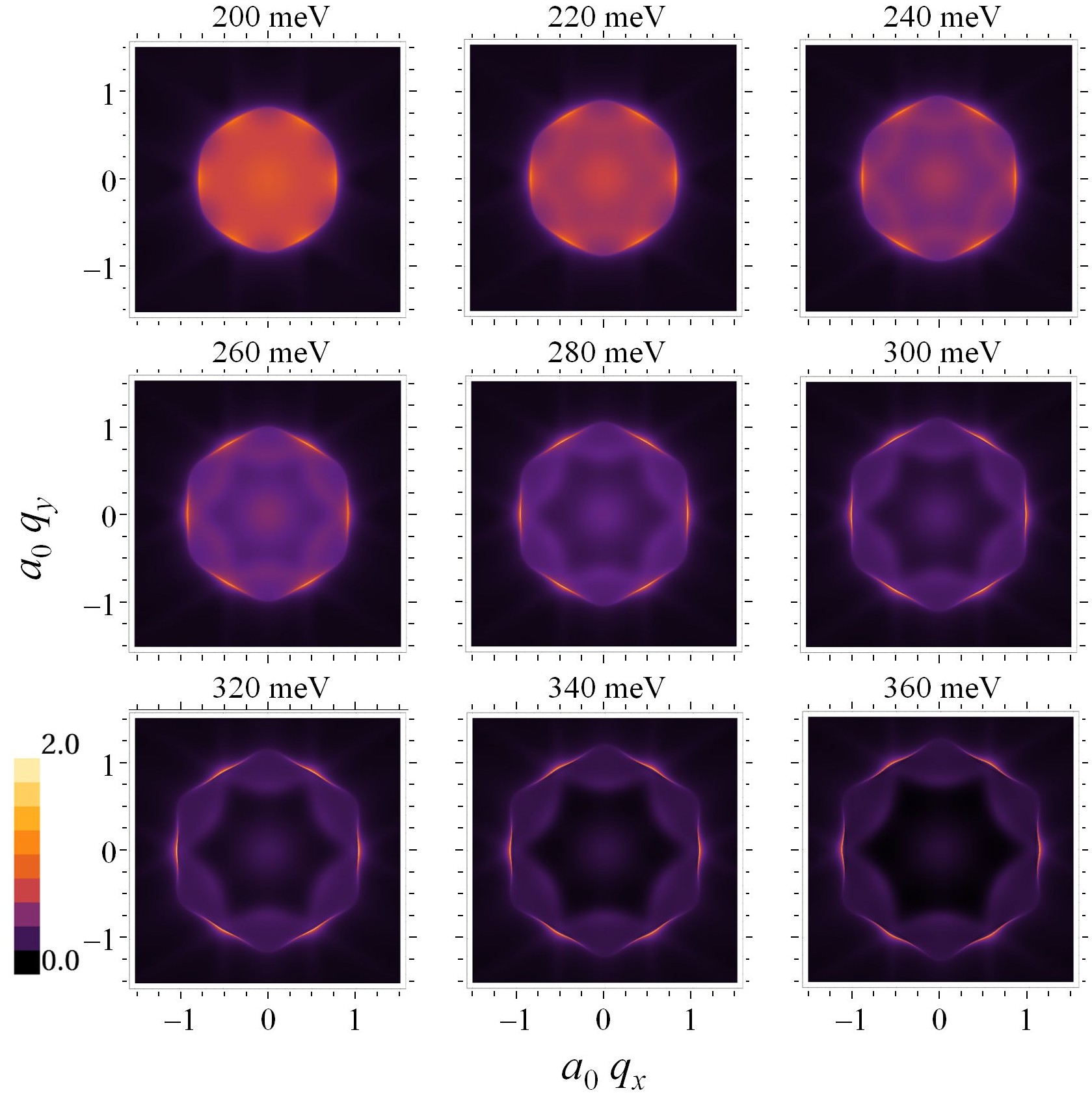}
\caption{(Color online)
QPI patterns $\Delta \rho({\bf{q}},\omega)$ at different energies $\omega$
for a dynamic magnetic impurity, with parameters
$A^2=2.23$, $\mu=100$\,meV, $U/v_F=0.3$, $\epsilon_d/v_F=-0.1$, and
$g/v_F=0.168$.
}
\label{fig:qpisc}
\end{figure}

The QPI pattern obtained at high energies 200--360\,meV for $\mu=100$\,meV
is shown in Fig.~\ref{fig:qpisc}. Upon increasing the scanning energy, the
high-intensity peaks move outwards and become more pronounced. This is
to be expected from the increasing diameter of the Fermi surface and the
increasing importance of hexagonal warping, which is due to the
underlying lattice structure and gives rise to the six-fold
symmetry.
We recall that our modelling neglects bulk bands which will contribute to the signal at
energies beyond the bulk gap.\cite{madhavan,heumen}

\begin{figure}[b]
\begin{center}
\includegraphics[width=0.47\textwidth]{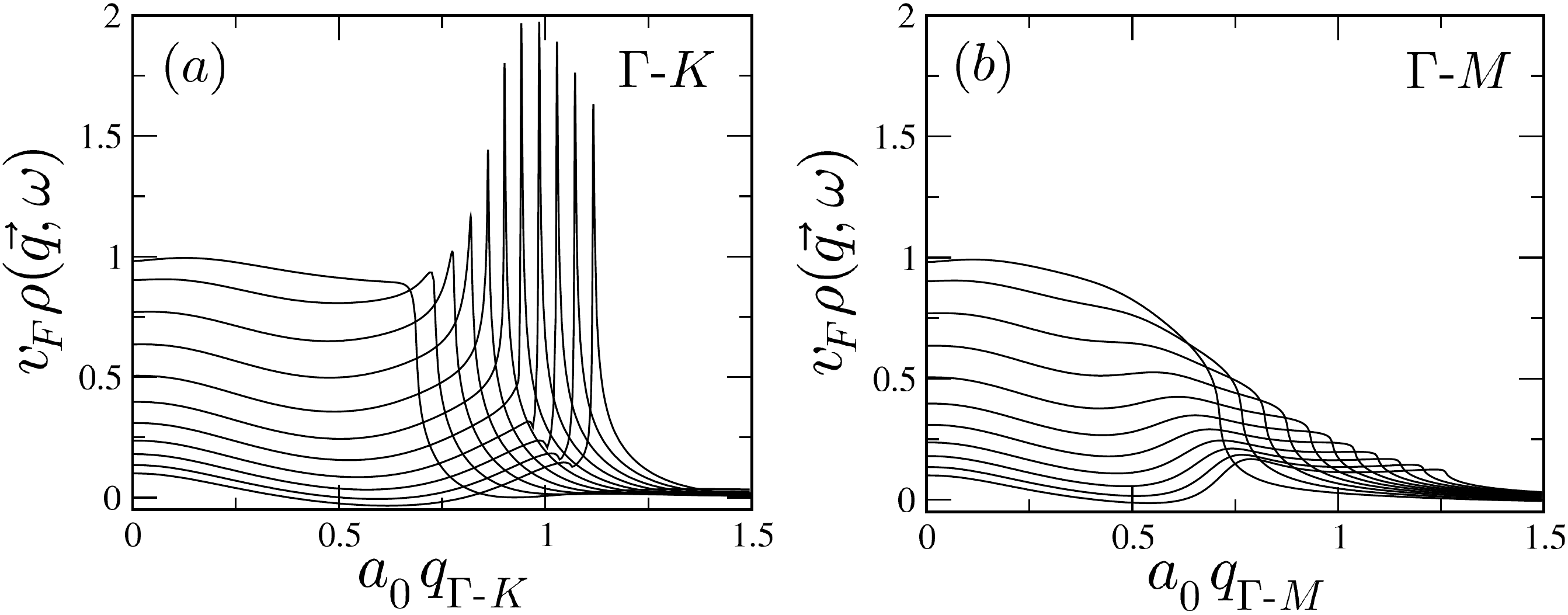}
\caption{\label{fig:cut_sc_hi}
Cuts through the QPI signal for a dynamic magnetic impurity along (a) the
$\Gamma$--$K$ direction, and (b) the $\Gamma$--$M$ direction, for a
system with the same parameters as in Fig.~\ref{fig:qpisc}. Energies
shown are from 160\,meV--360\,meV in steps of 20\,meV in order of increasing
peak position.
}
\end{center}
\end{figure}

Cuts through the QPI along the $\Gamma$--$K$ and $\Gamma$--$M$
directions shown in Fig.~\ref{fig:cut_sc_hi} allow a more detailed
analysis of the high-energy behavior.
One observes that the shape of the Hubbard satellites is manifest as a non-montonic
intensity variation of the QPI peaks with energy, in contrast to what would be observed
for non-magnetic impurities (see Appendix~\ref{sec:scalar}). Such a feature is
reminiscent of the experimental findings in Ref.~\onlinecite{madhavan}. We note that, in a
more complete modelling of the impurity, non-monotonic behavior could also arise from excited
crystal-field states of the magnetic impurity.

\subsubsection{Universal Kondo regime, $\omega\sim\TK$}\label{sec:universalregime}

\begin{figure}[t]
\centering
\includegraphics[width=0.49\textwidth]{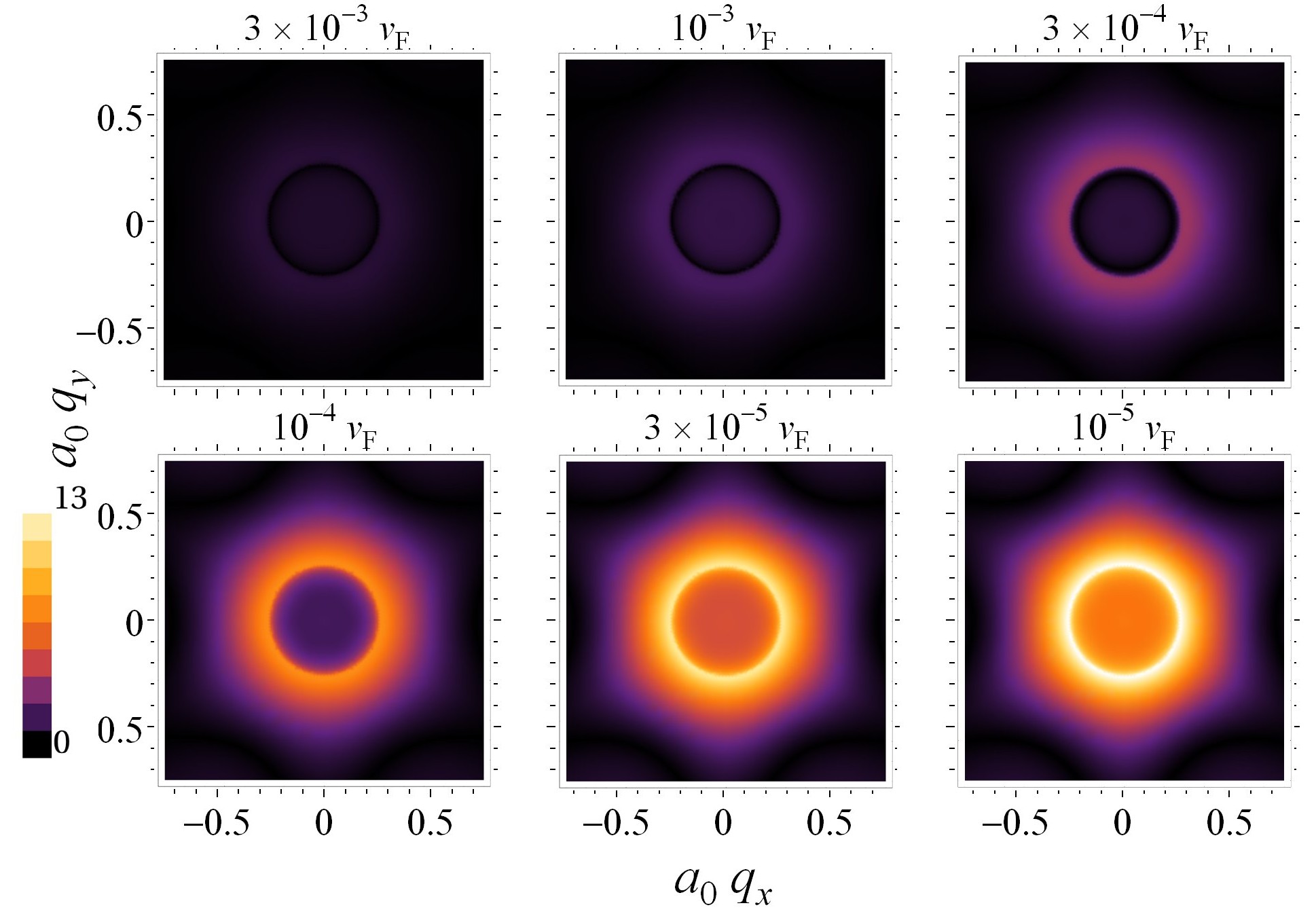}
\caption{(color online)
QPI patterns $|\Delta \rho({\bf{q}},\omega)|$ for a dynamic magnetic impurity in the
low-energy Kondo regime, with the same  parameters as Fig.~\ref{fig:qpisc}, but now with
scanning energies very close to the Fermi level.
Note the different scale as compared with Fig.~\ref{fig:qpisc};
also, $\Delta \rho({\bf{q}},\omega)$ has sign changes, see Fig.~\ref{fig:kondocut}.
}
\label{fig:kondoresonance}
\end{figure}

The pronounced build-up of impurity spectral intensity in a very narrow energy
window $\mathcal{O}(\TK)$ around the Fermi level is the characteristic
signature of the Kondo effect [see Fig.~\ref{fig:Gimp_LM}(b)]. This results in a
similarly characteristic evolution of the QPI pattern at low energies.

\begin{figure}[b]
\centering
\includegraphics[width=0.47\textwidth]{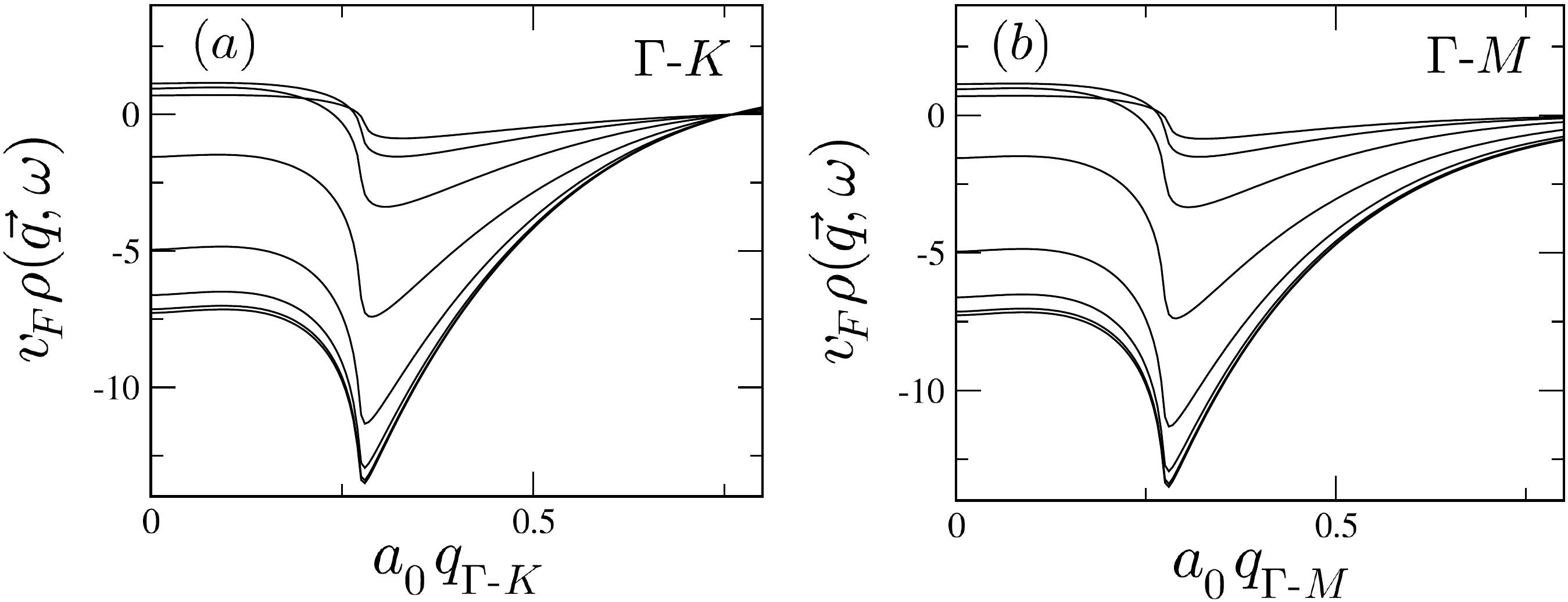}
\caption{
Cuts through the QPI signal for a dynamic magnetic impurity in the
low-energy Kondo regime along (a) the
$\Gamma$--$K$ direction, and (b) the $\Gamma$--$M$ direction, for a
system with the same parameters as in Fig.~\ref{fig:qpisc}. Energies shown are
$\omega/v_F=3 \times 10^{-3}$, $10^{-3}$, $3 \times 10^{-4}$, $10^{-4}$,
$3 \times 10^{-5}$, $10^{-5}$, $3 \times 10^{-6}$, $10^{-6}$ in order of increasing
peak intensity.
}\label{fig:kondocut}
\end{figure}

For the parameters used in Fig.~\ref{fig:qpisc}, the Kondo scale is
$\TK\approx 10^{-4}v_F$, and so we probe the system around these energies in
Fig.~\ref{fig:kondoresonance}. The intensity of the QPI peaks indeed
grows rapidly in this regime due to the Kondo effect. This is further
highlighted in Fig.~\ref{fig:kondocut}, where cuts through the QPI
pattern are shown. (We note the different scales in Figs.~\ref{fig:qpisc}
and~\ref{fig:kondoresonance} as well as Figs.~\ref{fig:cut_sc_hi}
and~\ref{fig:kondocut}, respectively.)

\begin{figure}[t]
\begin{center}
\includegraphics[width=0.47\textwidth]{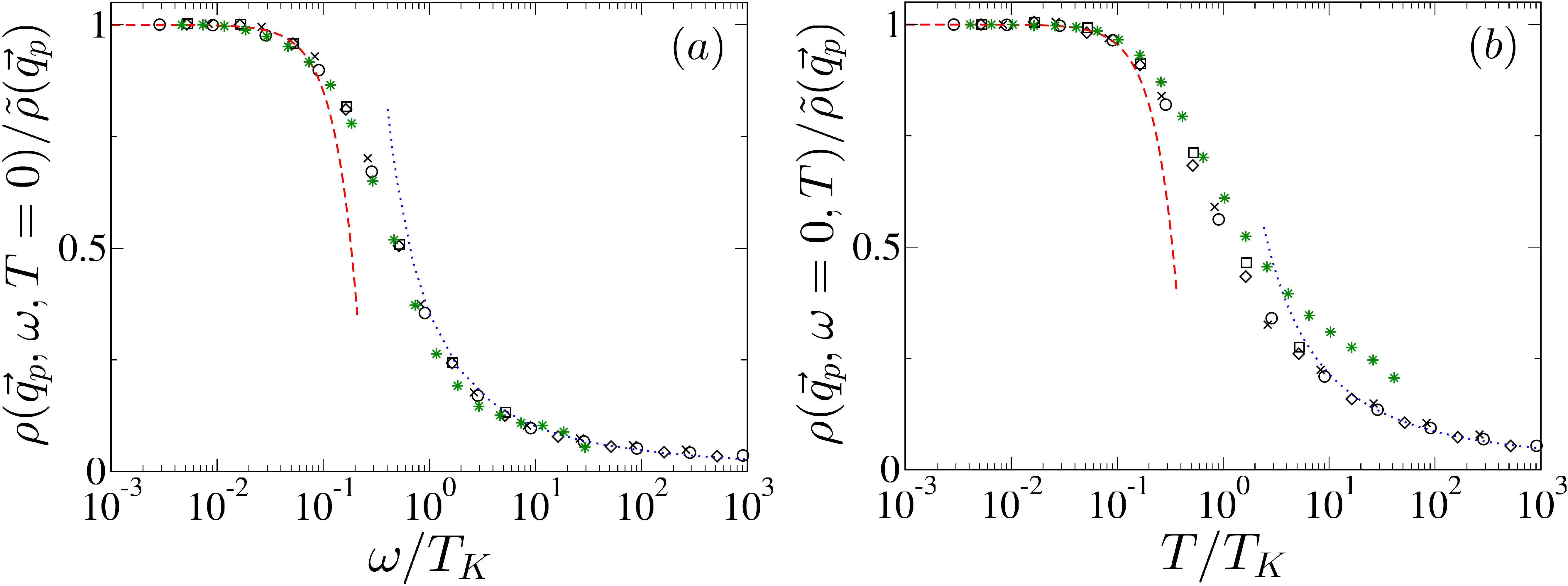}
\caption{\label{fig:universal}
(color online)
(a) Scaling collapse of the $T=0$ QPI peak intensity at low energies for
different bare parameters and hence different Kondo temperatures, $\TK$. The peak
position at $\textbf{q}=\textbf{q}_p$ is essentially pinned at low energies (see
Fig.~\ref{fig:kondocut}) as $\mu/v_F=0.1$ is kept fixed. Circles:
$(U/v_F,\epsilon/v_F,g/v_F)=(0.2,-0.1,0.126)$, $\TK\approx 10^{-2}$\,K; Squares:
$(0.2,-0.1,0.168)$, $\TK\approx 2$\,K; Diamonds: $(0.3,-0.15,0.126)$, $\TK\approx
6\times10^{-5}$\,K; Crosses:  $0.3,-0.15,0.168)$, $\TK\approx 0.1$\,K.
For illustration we have also plotted (Green Stars) the peak intensity for
$(U/v_F,\epsilon/v_F,g/v_F)=(0.2,-0.1,0.175)$ and a {\em different} $\mu/v_F=0.137$,
yielding $\TK\approx 20$\,K.
Red dashed line: Low-energy asymptote Eq.~\eqref{eq:asymptote1}. Blue dotted line: High-energy asymptote
Eq.~\eqref{eq:asymptote2}. (b) Scaling collapse of the $\omega=0$ peak intensity at low
temperatures for the same systems as in (a). Red dashed line: Low-temperature asymptote
Eq.~\eqref{eq:asymptote3}. Blue dotted line: High-temperature asymptote
Eq.~\eqref{eq:asymptote4}.
Deviations from scaling at higher $T$ become apparent for the $\TK=20$\,K data.
}
\end{center}
\end{figure}

At low energies in the Kondo regime, there is a universal scaling
collapse of the impurity Green function in terms of $\omega/T_K$ and $T/T_K$.
As in this regime (and in the scaling limit) the free Green function
$\hat{G}^{(0)}({\bf{k}},\omega)$ is essentially independent of
energy (and strictly independent of temperature) one naturally expects universality and
scaling of the \emph{entire} QPI pattern.
As we show in Fig.~\ref{fig:universal} this is indeed the case.

The $T=0$ peak intensity for different choices of the system parameters, and thus the
Kondo temperature, is shown in Fig.~\ref{fig:universal}(a). At low energies the data is
well described by
\begin{eqnarray}
\label{eq:asymptote1}
\frac{\rho({\bf{q}}_p,\omega,T=0)}{\tilde{\rho}({\bf{q}}_p)}&=&
1-a\left(\frac{\omega}{\TK}\right)^2,\quad \omega\ll\TK, \nonumber \\
{\rm{with}}\quad \tilde{\rho}({\bf{q}}_p)&=& \rho({\bf{q}}_p,\omega=0,T=0)
\end{eqnarray}
where ${\bf{q}}_p$ denotes the scattering vector of the peak. The quadratic behavior
reflects the Fermi liquid nature of the Kondo ground state. Similarly,
at elevated energies the asymptote is given by
\begin{equation}
\label{eq:asymptote2}
\frac{\rho({\bf{q}}_p,\omega,T=0)}{\tilde{\rho}({\bf{q}}_p)}
\sim\frac{1}{\ln(a'|\omega|/T_K)^2},\quad \TK\ll\omega\ll v_F,
\end{equation}
corresponding to spin-scattering processes in the vicinity of the local-moment fixed
point. These forms reflect universality in the scaling limit ($\TK\to 0$) of the Kondo
impurity, with universal coefficients $a,a'=\mathcal{O}(1)$.
Corrections arise from two sources: There is a leading linear-in-$\omega$ term in
Eq.~\eqref{eq:asymptote1} because the QPI pattern involves both real and imaginary parts
of $G^{d}_{\sigma}(\omega)$, and the real part of the self energy is generically
non-zero. The linear energy variation of $\rho^{(0)}$ contributes to this linear term as
well (this is where an explicit $\mu$ dependence enters), and it also causes deviations
in Eq.~\eqref{eq:asymptote2}. These corrections are suppressed by the factor
$\omega/{\rm min}(|\mu|,v_F)$ and are small for the data shown in Fig.~\ref{fig:universal}(a).

The temperature scaling of the $\omega=0$ peak intensity is shown
in Fig.~\ref{fig:universal}(b). In the low-temperature limit one finds
\begin{equation}
\label{eq:asymptote3}
\frac{\rho({\bf{q}}_p,\omega=0,T)}{\tilde{\rho}({\bf{q}}_p)}=
1-b\left(\frac{T}{\TK}\right)^2,\quad T \ll \TK,
\end{equation}
characteristic of Fermi-liquid behavior. Indeed, we find that the Fermi liquid relation $a/b=3/\pi^2$ is well-satisfied from our numerical calculations. 
Similarly to Eq.~\eqref{eq:asymptote2}
the spin-flip scattering processes in the vicinity of the local-moment fixed point
lead to the temperature dependence
\begin{equation}
\label{eq:asymptote4}
\frac{\rho({\bf{q}}_p,\omega=0,T)}{\tilde{\rho}({\bf{q}}_p)}\sim
\frac{1}{\ln(b' T/\TK)^2},\quad \TK \ll T \ll v_F,
\end{equation}
where $b,b'=\mathcal{O}(1)$ are again universal. Deviations from universality arise as
above; they are visible for the highest-$\TK$ data in Fig.~\ref{fig:universal}(b).

The characteristic feature of the Kondo effect is thus the rapid increase of QPI peak
intensity as the scanning energy approaches the Fermi level, together with the
(approximate) scaling collapse in $\omega/\TK$ and $T/\TK$.
These features should be readily observable provided that experiments are performed
at temperatures of order $\TK$ or lower.

Finally, we comment briefly on the case where a (small) magnetic field acts on the
dynamic magnetic impurity. Since $T^0_{\textbf{k},\textbf{k}'}(\omega)$ and the impurity
model in the absence of a field is SU(2) symmetric, it is sufficient to discuss the
spin-summed impurity spectral function. For small fields $\mathcal{O}(T_K)$, this is
known to develop a split Kondo resonance,\cite{hewson} signatures of which should show up
in QPI. For larger fields $\gg T_K$, the Kondo effect is destroyed entirely, and only the
high-energy Hubbard satellites remain thus leading to a behavior qualitatively similar to
Fig.~\ref{fig:qpisc}.

\subsection{Local-moment phase}\label{lmphase}

In the local-moment regime, realized if the chemical potential is tuned to the Dirac
point $\mu=0$ (unless the Kondo coupling and asymmetry is very strong), there is no low-energy Kondo
resonance in the impurity spectral function, although the Hubbard satellites remain [see
Fig.~\ref{fig:Gimp_LM}(a)].

As a result, the high-energy features of the QPI pattern are essentially the same as
those in Figs.~\ref{fig:qpisc}, \ref{fig:cut_sc_hi}, but there is no build-up of QPI peak
intensity in the vicinity of the Fermi level. The same applies to the situation with non-zero
$\mu$ but temperatures $T\gg\TK$.


\section{Conclusions}\label{sec:conclusion}

In this work we have studied modulations in the LDOS caused by dilute magnetic impurities
on the surface of 3D TIs. Despite the coupling of orbital and spin degrees of freedom in
the helical surface metal and hexagonal warping due to the underlying TI lattice
structure, the quantum impurity problem itself is of conventional type, such that Kondo
screening is generically present in the low-temperature limit (unless the chemical
potential is tuned to the Dirac point).

We identified the {\em energy} dependence of the QPI peak intensity as the fingerprint of
dynamic magnetic impurities (as opposed to non-magnetic or spin-polarized impurites, see
appendix). At elevated energies, non-monotonic QPI peak intensity for scanning energies
is observed due to Hubbard satellites in the impurity spectral function. At low scanning
energies, Kondo screening of the impurity produces a strong build-up of QPI peak
intensity, whose energy and temperature dependence are universal functions of
$\omega/\TK$ and $T/\TK$.

However, the {\em momentum} dependence of the QPI signal (at fixed energy) is identical
for the different types of impurities, due to the fact that only the spin-diagonal part
of the T matrix produces modulations in the charge channel.\cite{guofranz}
This casts doubts on the interpretation of the experimental data given in
Ref.~\onlinecite{madhavan}, where the magnetic character of dilute Fe impurities was made
responsible for the appearance of new QPI wavevectors.
At present, the source of this QPI signal is unclear, and more systematic studies are
called for.


\acknowledgments

We acknowledge useful discussions with R. Bulla, M. Golden, E. van Heumen, J. Paaske, A.
Rosch, and E. Sela. This work was supported by the German Research Foundation (DFG)
through the Emmy-Noether program under FR 2627/3-1 (LF) and SCHU 2333/2-1 (DS) as well as
SFB 608 (AKM,LF), FOR 960 (AKM,MV), and GRK 1621 (MV).


\appendix
\section{Local Green function and symmetries}\label{app:localgreen}

Here we consider the form of the local free Green function, and show
explicitly that its off-diagonal components vanish. The inverse Green
function for Matsubara frequencies in spin-space reads
\begin{widetext}
\begin{equation}
\hat{G}^0(i\omega_n,{\bf{k}})^{-1}=  \left( \begin{array}{cc}   i \omega_n+\mu-A^2 k^3 \cos 3 \phi_{\bf{k}} & -v_F \left(k_y+ik_x \right)\\ -v_F \left(k_y-ik_x \right)& i \omega_n+\mu+A^2 k^3 \cos 3 \phi_{\bf{k}}\end{array} \right)
\end{equation}
where the local Green function assumes the form
\begin{equation}
\hat{G}^0(i\omega_n,r=0)=\int \frac{kdk}{2\pi}\int\frac{d\phi}{2\pi} \frac{1}{(i\omega_n+\mu)^2-v_F^2(k^2+A^4 k^6 \cos^2 3\phi)}\left( \begin{array}{cc}   i \omega_n+\mu-A^2 k^3 \cos 3 \phi & v_F \left(k \sin \phi +ik \cos \phi \right)\\ v_F \left(k \sin \phi-ik \cos \phi \right)& i \omega_n+\mu+A^2 k^3 \cos 3 \phi\end{array} \right)
\end{equation}
The off-diagonal terms do not survive
the angular integration, such that the remaining diagonal components simplify to
\begin{equation}
\hat{G}^0(i\omega_n,r=0)= \mathbb{I} \int \frac{kdk}{2\pi}\int\frac{d\phi}{2\pi} \frac{i\omega_n+\mu}{(i\omega_n+\mu)^2-v_F^2(k^2+A^4 k^6 \cos^2 3\phi)}
\end{equation}
\end{widetext}
This immediately allows one to read off the energy eigenvalues in Eq.~\eqref{eq:spectrum}.


\section{QPI from non-magnetic impurities}
\label{sec:scalar}

For completeness, we show here the QPI signal of non-magnetic impurities, which has
largely been calculated and discussed in Refs.~\onlinecite{yazdani,guofranz,zhang,madhavan,zhou09}.

\subsection{Scattering potential}

The TI surface metal with a non-magnetic scalar impurity or potential scatterer is described by
$H=H_0+H_V$, with
\begin{eqnarray}\label{eq:imp}
H_V=V \sum_\sigma \Psi^\dagger_{\sigma}(r=0)\Psi^{\phantom{\dagger}}_{\sigma}(r=0)\;,
\end{eqnarray}
The exact T matrix is
\begin{eqnarray}\label{eq:tmps}
\hat{T}_{\textbf{k},\textbf{k}'}(\omega)=\frac{1}{N}\frac{V}{1-V f_{\omega,\mu}}\mathbb{I}
=\frac{1}{N}T^0(\omega)\,\mathbb{I}\;,
\end{eqnarray}
with $f_{\omega,\mu}$ given in Eq.~(\ref{eq:lg}), while lowest-order Born
approximation (valid for small $V$) corresponds to $T^0(\omega)=V$
which is real and constant. The QPI pattern in this approximation
is displayed in Fig.~\ref{fig:dos}(b).

\begin{figure}[t]
\begin{center}
\includegraphics[width=0.47\textwidth]{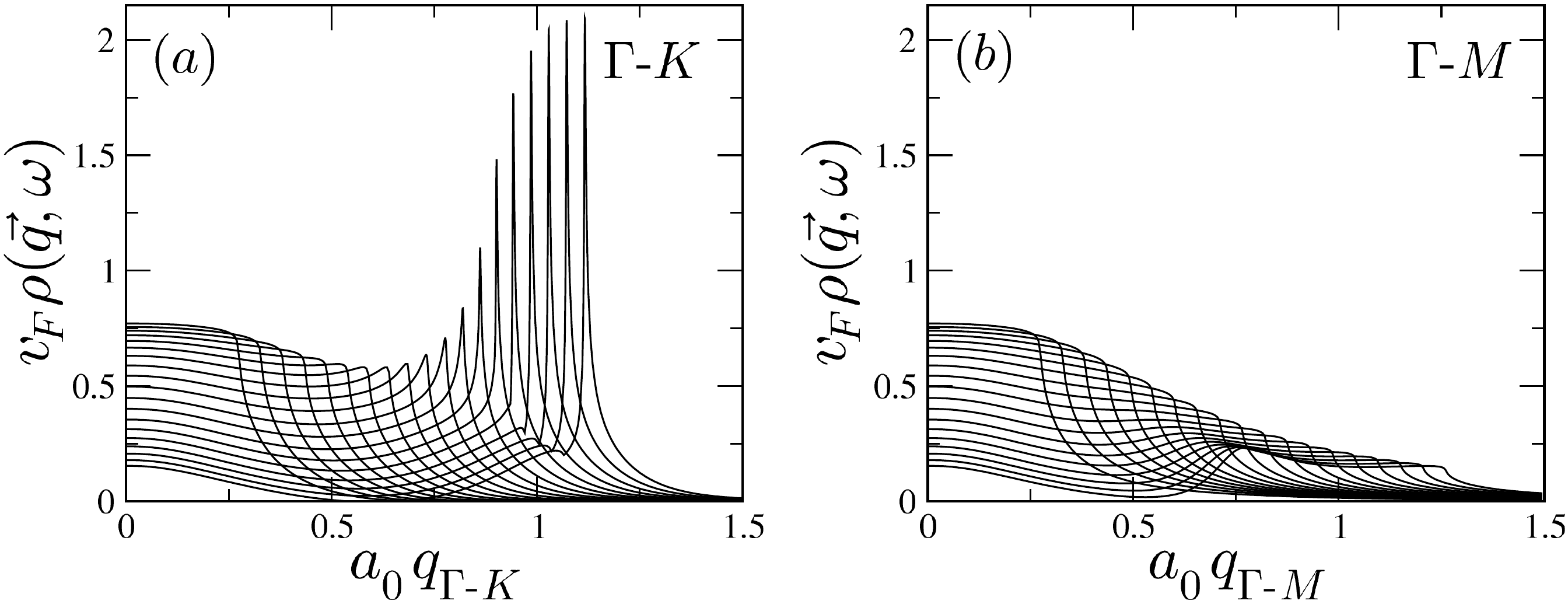}
\caption{\label{fig:cut_ps}
Cuts through the QPI signal for a scalar impurity along (a) the
$\Gamma$--$K$ direction, and (b) the $\Gamma$--$M$ direction, for a
system with $\mu=100$\,meV and a potential scattering strength
$V/v_F=0.1$. Energies shown are from 0\,meV--360\,meV in steps of 20meV
in order of increasing peak position (20\,meV $\simeq 0.027v_F$).
}
\end{center}
\end{figure}

\begin{figure}[b]
\begin{center}
\includegraphics[width=0.47\textwidth]{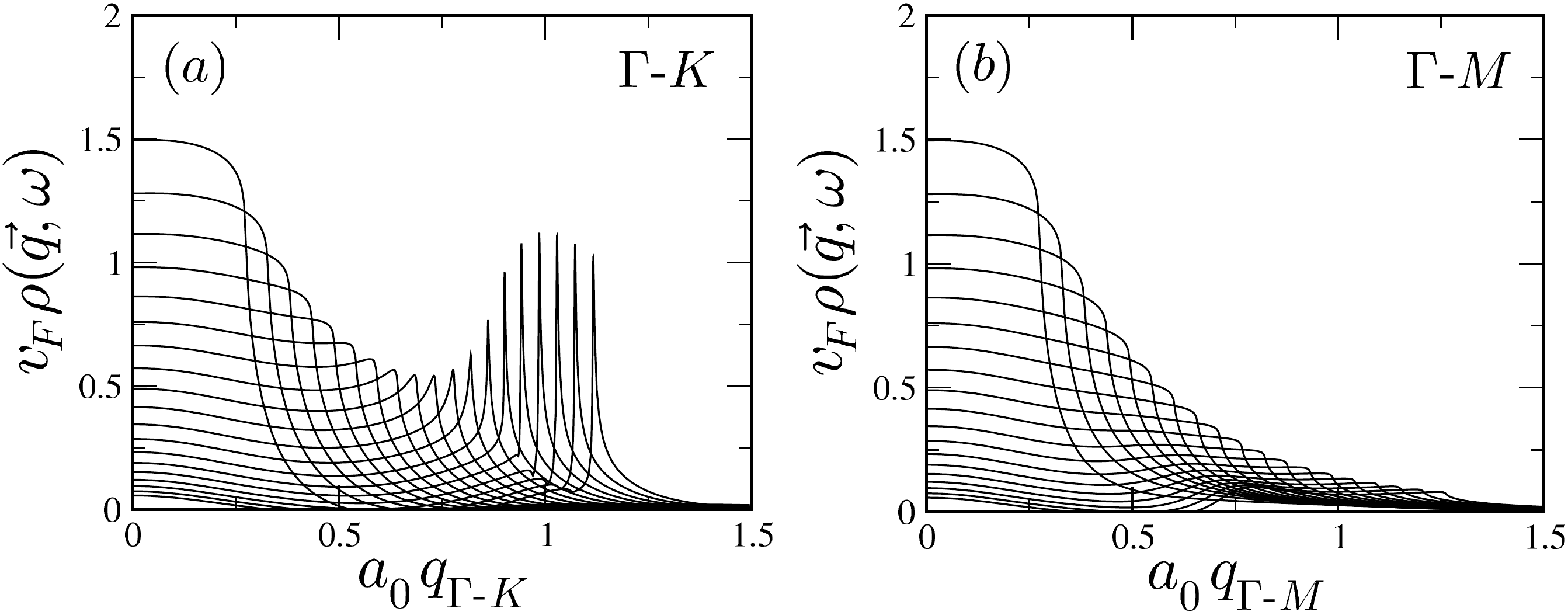}
\caption{\label{fig:cut_u0}
Cuts through the QPI signal for a resonant level along (a) the
$\Gamma$--$K$ direction, and (b) the $\Gamma$--$M$ direction, for a
system with $\mu=100$\,meV, $U=0$, $\epsilon_d/v_F=-0.1$ and
$g/v_F=0.168$. Energies shown are from 0\,meV--360\,meV in steps of 20\,meV
in order of increasing peak position.}
\end{center}
\end{figure}

Since impurities are often not weak, the Born approximation may be insufficient.
Results using the full T matrix, which now includes a small imaginary part,
are plotted in Fig.~\ref{fig:cut_ps} for $V/v_F=0.1$. We observe that the
intensity of the peak along the $\Gamma$--$K$ direction simply increases
monotonically as the scanning energy $|\omega |$ increases (for both
$\omega>0$ and $\omega<0$).

\subsection{Resonant level}\label{sec:QPIRL}

A so-called resonant level impurity is obtained by setting $U=0$ in Eq.~(\ref{Eq:Himp}),
physically corresponding to a mixed-valent impurity atom. Then, the impurity spectral
function consists of a \emph{single} peak centered around $\omega=\epsilon_d$. In
contrast to the scalar impurity discussed above the T matrix of the resonant level
possesses an appreciable imaginary part (see Eq.~\eqref{eq:gfaim} with $\Sigma^d=0$).

Cuts through the QPI pattern along the $\Gamma$--$K$ and $\Gamma$--$M$
directions at different scanning energies are plotted in
Fig.~\ref{fig:cut_u0} for $\mu=100$\,meV, $\epsilon_d/v_F=-0.1$
and $g/v_F=0.168$. For this resonant level, the intensity of the peak is reminiscent of the high-energy response produced by a Kondo impurity obtained for $U>0$ (simple charge fluctuations are responsible for the evolution of the QPI signal at high energies in both cases). However, crucially there is no build-up of QPI intensity at \emph{low} energies, since there is no Kondo effect (see by contrast Fig.~\ref{fig:kondocut}).


\section{QPI from static magnetic impurities}\label{sec:staticmag}

The simplest model for a static, i.e. polarized, magnetic impurity is a local magnetic
field, corresponds to a spin-dependent version of the potential scattering
case considered above. Thus we have $H=H_0+H_h$, with
\begin{eqnarray}\label{eq:smi}
H_h=h \sum_\sigma \Psi^\dagger_{\sigma}(r=0)\sigma_z\Psi^{\phantom{\dagger}}_{\sigma}(r=0)\;.
\end{eqnarray}
Within the Born approximation, the T matrix is then
\begin{eqnarray}
\hat{T}_{{\bf{k}},{\bf{k}}'} \approx \frac{h}{N}\sigma_z\;,
\end{eqnarray}
i.e. $T^1(\omega)=h$, such that there is {\em no} response in QPI due to
Eq.~(\ref{eq:qpidef}). Beyond the lowest-order Born approximation,
a (weak) response similar to that of a potential scatterer is induced, see
Refs.~\onlinecite{xu,guofranz,zhou09}.



\begin{thebibliography}{43}
\bibitem[{con()}]{cont_note}
\bibinfo{note}{Recall that our model is defined in the continuum; the
  short-distance (or ultraviolet) cutoff is effectively set by $a_0$.}
\end{thebibliography}
\end{document}